
%
%
%
%
\input phyzzx.tex
\pubnum={DPNU-95-26}
\date={August 1995}
\titlepage
\title{Charge Condensation in QED$_3$ with a Chern-Simons Term}
\vskip 1.5cm
\author{Taichi ITOH}
\address{Department of Physics, Nagoya University\break
        Nagoya, Japan, 464-01}
\andauthor{Toshiro SATO}
\address{Matsusaka University\break
        Matsusaka, Mie, Japan, 515}
\vskip 2cm
\abstract{ Introducing a chemical potential in the functional method, we
construct the effective action of QED$_3$ with a Chern-Simons term. We
examine a possibility that charge condensation $\langle\psi^\dagger\psi
\rangle$ remains
nonzero at the limit of the zero chemical potential. If it happens,
spontaneous magnetization occurs due to the Gauss' law constraint which
connects the charge condensation to the background magnetic field. It is
found that the stable vacuum with nonzero charge condensation is realized
only when fermion masses are sent to zero, keeping it lower than the
chemical potential. This result suggests that the spontaneous magnetization
is closely related to the fermion mass.}
\vfill
\eject
Field theoretical models in 2+1 dimensional space-time have been investigated
from various interests. In particular, models coupled to the Chern-Simons
gauge field in 2+1 dimensional space-time have attracted
much attention, since those are known to be strongly associated with the
fractional quantum Hall effect and the high $T_c$ superconductivity.$^{[1,2]}$
Furthermore, it was shown that the Lorentz invariance should be
``spontaneously" broken in a model in which fermions are coupled to a
Maxwell-Chern-Simons gauge field.$^{[3]}$

The meaning of ``spontaneous" breakdown of the Lorentz invariance in those
works is that there exists a state with nonzero magnetic field, $B\ne 0$, and
with the energy lower than the ``lowest"
energy state in the $B=0$ states. So, it is expected that the real vacuum
would exist in $B\ne 0$ states.

However, the above discussion is not sufficient to confirm
that the Lorentz invariance is really broken ``spontaneously".
One way to confirm that the Lorentz invariance is broken ``spontaneously" is
as follows; first of all,
we should introduce a source term for the order parameter characterizing the
breakdown, and calculate vacuum expectation value of the order parameter.
Next, we have to show that the vacuum expectation value remains nonzero when
the source approaches  zero.

In this paper, we shall re-examine the model from the above view point, and
investigate the vacuum structure of the model. Moreover, we
shall find out what kind of criterion is needed to give rise to
breakdown of the Lorentz invariance.

To do this, it is convenient to adopt the Euclidean formulation of the
quantum field theory. In this case, Lorentz symmetry is shifted to the
$SO(3)$ symmetry in the Euclidean space-time. So, the
dynamical generation of the magnetic field means spontaneous breakdown of
the $SO(3)$ symmetry. In the following, we will mainly
discuss whether or not the magnetic field is dynamically generated, when we
take a limit where the $SO(3)$ symmetry in the effective action is restored.
Lagrangian of the system in the Euclidean formulation
is given as
$$
{\cal L}=
{1\over 4}F_{\mu\nu}F_{\mu\nu}
-i{1\over 2}\kappa\epsilon_{\mu\nu\rho}A_\mu\partial_\nu A_\rho
+\sum_a\{ -{\bar\psi}_a[\gamma^a_\mu(\partial_\mu-iq_aA_\mu)+m_a]\psi_a\}\ ,
\eqno(1)
$$
where fermions are two-component Dirac spinors, and $a$ denotes the fermion
flavor number. Since the following analysis is irrelevant to the flavor
number, discussion is restricted to one flavor case. We
note that in this system the magnetic field $B$ is generated by the fermion
density $\rho\equiv\langle\psi^\dagger\psi\rangle$, since the Gauss' law in
this model is represented as
$$
\rho={\kappa\over q}B
\eqno(2)
$$
Namely, nonzero vacuum expectation value of $B$ is due to fermion
charge condensation. This fact is a crutial point in the following discussion.

In the ref.[3], it was assumed from the begining that there exists a stable
vacuum at $B\ne 0$, i.e. at finite fermion density. In those works, the vacuum
state with
finite fermion density was realized by introducing the filling factor $\nu$ of
Landau levels. Then, they calculated the energy difference $\Delta E=E(B\ne 0)
-E(B=0)$ by the use of fermion operator $\psi$ expanded at each vacuum,
taking account of quantum fluctuations. However, $\nu$ represents
occupation rate of fermion only on a certain Landau level. If we introduce a
filling factor for a certain Landau level into the theory, we can deal with
dynamics only on a Landau level associated with the introduced filling
factor. So, we have to examine dynamics on every Landau level, separately.
This means that we cannot deal with the model from the global view point,
in which all of the Landau levels are controlled by a single continuous
parameter.

Alternatively, we can also represent finite density state by the use of
chemical potential $\mu$. In this case, as we will show below, $\mu$ is a free
parameter of the theory, and can express states for all of the Landau
levels. So, we can examine configuration of vacua of the model, by varing the
value of $\mu$.

Now, let us examine vacuum structure of the model. As mentioned above, we have
to add a chemical
potential term in the action in order to take account of $B\ne 0$, since $B
\ne 0$ is associated with nonzero fermion density because of the Gauss'
law.$^{[2]}$
Adding the chemical potential term and integrating out the fermion field, we
obtain the action for the gauge field $A_\mu$
$$
\eqalign{
S_{\rm eff}[A_\mu;\mu,m]&=\int d^3x\ {1\over 4}F_{\mu\nu}F_{\mu\nu}
-i{1\over 2}\kappa\epsilon_{\mu\nu\rho}A_\mu\partial_\nu A_\rho\cr
&\qquad\qquad-{\rm tr}\ln[\gamma_\mu D_\mu[A_\mu]+\mu\gamma_3+m]\ .\cr
}
\eqno(3)
$$
Next, separating the gauge field into two parts as
$$
A_\mu\equiv A^{\rm c}_\mu+{\tilde A}_\mu\ ,
$$
where $A^{\rm c}_\mu$ is a classical background field part
and ${\tilde A}_\mu$ is a quantum
fluctuation part, we can regard $S_{\rm eff}[A^{\rm c}_\mu+{\tilde A}_\mu;\mu,
m]$ as a classical action for ${\tilde A}_\mu$. Then, integrating out ${
\tilde A}_\mu$, we can obtain the effective action for the classical gauge
field, $\Gamma_{\rm eff}[A^{\rm c}_\mu;\mu,m]$, which is defined as$^{[4]}$
$$
e^{-\Gamma_{\rm eff}[A^{\rm c};\mu,m]}
=\int[d{\tilde A}_\mu]\ \exp\Bigl[-(S_{\rm eff}[{\tilde A}_\mu
+A^{\rm c}_\mu;\mu,m]
-\int d^3x\ {\delta\Gamma_{\rm eff}\over\delta{\tilde A}_\mu}A^{\rm c}_\mu)
\Bigr]\ .
\eqno(4)
$$
At the one loop level, $\Gamma_{\rm eff}[A^{\rm c}_\mu;\mu,m]$ is
expressed as
$$
\eqalign{
\Gamma_{\rm eff}[A^{\rm c}_\mu;\mu,m]&=S_{\rm eff}[A^{\rm c}_\mu;\mu,m]\cr
&\quad+{1\over 2}\ln{\rm det}[\Delta_{\mu\nu}^{-1}(-i\partial)-\Pi_{\mu\nu}
(-i\partial;\mu,m,A^{\rm c}_\mu)]\cr
&\quad+({\rm higher\ loop\ corrections})\ ,\cr
}
\eqno(5)
$$
where $\Delta_{\mu\nu}$ denotes the bare photon propagator and $\Pi_{\mu\nu}$
is the vacuum polarization tensor at the one loop level of
fermion coupled to the background field $A^{\rm c}_\mu$.
Here, since vacuum must be a stationary point, the following condition
is satisfied;
$$
{\delta\Gamma_{\rm eff}[A^{\rm c}_\mu;\mu,m]\over\delta A^{\rm c}_\mu}=0\ .
\eqno(6)
$$
Since this is just equivalent to the Gauss' law, solution of eq.(6),
$A^{\rm c\ sol}_\mu$, satisfies eq.(2).

Generally, it is known that the chemical potential in free fermion system
gives the Fermi energy. In the case of eq.(3), it is found that the
chemical potential gives the highest energy values of occupied Landau levels.
In this sense,
it can also be said that the chemical potential is indirectly associated with
the filling factor $\nu$ of Landau levels. Indeed, value of the chemical
potential (i.e. the highest Landau level energy) can be
determined by the relation
$$
\rho[\mu,A^{\rm c}_\mu]
={\partial\Gamma_{\rm eff}[A^{\rm c}_\mu;\mu,m]\over\partial\mu}\ ,
\eqno(7)
$$
for a fixed value of charge density. It is also found,
however, that the chemical potential $\mu$ can be regarded as a source of
order parameter for the charge condensation (or, equivalently, order parameter
for the magnetization), when we consider the chemical potential as a
free parameter. This fact implies that if we calculate the value of the right
hand side of eq.(7) in the limit of $\mu\rightarrow 0$, we can make sure
of possibility of the ``spontaneous" charge condensation or magnetization;
that is, if the value of $\partial \Gamma_{\rm eff}[A^{\rm c}_\mu;\mu,m]/
\partial\mu|_{A^{\rm c}\equiv A^{\rm c\ sol}}$ at the limit ${\mu\rightarrow
0}$ remains nonzero, it can be said that there is a candidate for a stable
vacuum state at $\rho\ne 0$ (i.e. at $B\ne 0$), and the
``spontaneous" charge condensation or $B$ generation can actually occur. Here,
we have to notice that the solution $A^{\rm c\ sol}_\mu
=-B\delta_{\mu 1}x_2$, which
is substituted into the right hand side of eq.(7), is strongly related to
the charge density $\rho$ in the left hand side on the vacuum, via the
Gauss' law.

First, we neglect, for simplicity, the quantum fluctuation part of the photon.
In this case,
it is easy to differentiate the effective action $\Gamma_{\rm eff}[A^{\rm
c}_\mu;\mu,m]$ with respect to $\mu$. By using the fact that explicit
form of the fermion propagator under the external magnetic field $B$ is
expressed as$^{[5]}$
$$
S_F(x,y;\mu)={\rm exp}\bigl( iq\int^x_yd\xi_\mu A^{\rm c\ sol}_\mu(\xi)\bigr)
{\tilde S}_F(x-y;\mu)\ ,
\eqno(8)
$$
where
$$
{\tilde S}_F(x-y;\mu)\equiv\int{d^3k\over(2\pi)^3}S_F(k')e^{ik(x-y)}\ ,
$$
and
$$
\eqalign{
S_F(k)\equiv&\int^\infty_0ds\ \exp\Bigl\{-s\Bigl[m^2+k_3^2+{\vec k}^2
{\tanh(qBs)\over qBs}\Bigr]\Bigr\}\cr
&\qquad\qquad\times(-i\gamma_\mu k_\mu+m)
(1+\gamma_3{qB\over|qB|}\tanh(qBs))\ ,\cr
}
\eqno(9)
$$
with $k'_\mu\equiv k_\mu-i\mu\delta_{\mu 3}$, we obtain$^{[2]}$
$$
\eqalign{
\rho[\mu,A^{\rm c\ sol}_\mu]
&=\int{d^3k\over(2\pi)^3}{\rm Tr}[\gamma_3S_F(k')]\cr
&={|qB|\over 2\pi}
\Bigl[{\rm Int}\Bigl({\mu^2-m^2\over 2|qB|}\Bigr)
+{1\over 2}\Bigr]\theta(\mu-|m|)\cr
&\qquad+{qB\over 4\pi}{m\over|m|}\theta(|m|-\mu)+{\cal O}(q^3)\ .\cr
}
\eqno(10)
$$
In eq.(10), we assumed that $\mu>0$. Then, we have only to see the
value of $\rho$ in eq.(10) in the limit
of $\mu\rightarrow 0$. As mentioned before, however, at this stage, we should
substitute the
Gauss' law into eq.(10) to remove $B$ or $\rho$ in the expression, so
that eq.(10) is expressed only by $\rho$ or $B$. For example, let us remove
$\rho$. Thus, eq.(10) is rewritten as
$$
B\Bigl\{\kappa-{q^2\over 2\pi}{qB\over|qB|}\Bigl[{\rm Int}
\Bigl({\mu^2-m^2\over 2|qB|}\Bigr)+{1\over 2}\Bigr]\theta(\mu-|m|)
-{q^2\over 4\pi}{m\over|m|}\theta(|m|-\mu)\Bigr\}=0\ ,
\eqno(11)
$$
where the higher order terms ${\cal O}(q^3)$ are neglected.
Therefore, if $B\ne 0$ at vacuum state, then equation
$$
\kappa-{q^2\over 2\pi}{qB\over|qB|}\Bigl[{\rm Int}
\Bigl({\mu^2-m^2\over 2|qB|}\Bigr)+{1\over 2}\Bigr]\theta(\mu-|m|)
-{q^2\over 4\pi}{m\over|m|}\theta(|m|-\mu)=0\
\eqno(12)
$$
must be satisfied. We note that the left hand side of eq.(12)
is not an analytic function, since there are singularities at $\mu^2=
m^2+2n|qB|\quad(n=0,1,2,\cdots)$ which correspond just to energy values of
the Landau levels (vertical line in Fig.1). At
those points, the analysis might fail to hold  ( in order to remove the
singularities from the discussion, we should extend the system to finite
temperature case). Therefore, in the following discussion, we concentrate to
the analytic region (horizontal line in Fig.1).

Now, we will see whether eq.(12) actually holds, when the limit of $\mu
\rightarrow 0$ is taken. To see this, it is convenient to regard the fermion
mass as
a free parameter in the theory as well as the chemical potential $\mu$, so
that ($\mu,m$)-plane can be considered. Let us divide the plane
into two regions; $|m|>\mu$ and $|m|<\mu$. In the region of $|m|>\mu$, only if
$\kappa$ satisfies
$$
\kappa={ q^2\over 4\pi}{m\over|m|}\ ,
\eqno(13a)
$$
eq.(12) holds, independent of the value of $\mu$. In other words, there
is a possibility that nontrivial vacuum with $B\ne 0$ could
exist, when eq.(13a) holds.
On the other hand, in the region $|m|<\mu$, eq.(12) has solutions when $\kappa$
satisfies
$$
\kappa={qB\over|qB|}{q^2\over 4\pi}(2n+1)\qquad(n=0,1,2\cdots)\ .
\eqno(13b)
$$
Indeed, the solution of eq.(12) is given by the following inequality;
$$
m^2+2n|qB|<\mu^2<m^2+2(n+1)|qB|\ .
\eqno(14)
$$
One can find from eq.(14) that if we choose $n=0$ (i.e., only the lowest
Landau level is filled) and take $m\rightarrow 0$ limit, then $\mu
\rightarrow 0$ limit can be taken so that eq.(12) holds. Therefore, if
these conditions are satisfied, a nontrivial vacuum with $B\ne 0$ could
exist even in the region $|m|<\mu$.

Next, let us consider whether the situation could be changed, if quantum
fluctuation terms of the photon and the higher order correction terms in
eq.(10) are taken into account in the effective action. Some years
ago, by extending the nonrenormalization theorem of Coleman-Hill$^{[6]}$ in
the zero-density case to the finite density case, it was shown by Lykken et
al.$^{[2]}$ that in the system of a
Chern-Simons  gauge field coupled to fermions, the odd parity part of vacuum
polarization tensor $\Pi_{\rm odd}(0)$ is unaffected by the higher order
radiative corrections.
Since charge density is associated with $\Pi_{\rm odd}$ through the Gauss'
law, it is also unaffected by the higher order corrections. There does not
exist the Maxwell term in their model. However, it is expected that existence
of the Maxwell term does not play an essential role in their conclusions. Then,
their
conclusions are still valid for our case, which has the Maxwell term.
The above result therefore holds exactly even if the quantum effects are taken
into account.

It is not yet clear whether or not the candidates for the vacuum state
obtained through the above analysis are actually the stable vacua. In order
to clarify this
point, we have to calculate energy difference $\Delta E=E(B\ne 0)-E(B=0)=
(\Gamma_{\rm eff}[A^{\rm c}=A^{\rm c\ sol}]-\Gamma_{\rm eff}[A^{\rm c}=0])
/$(space-time volume).
After some calculations, the energy difference is expressed as
$$
\eqalign{
\Delta E&={1\over 2}\int{d^3p\over(2\pi)^3}
\ln{{\rm det}D^{-1}(p;B)\over{\rm det}D^{-1}(p;B\rightarrow 0)}
+\Delta E_F+{B^2\over 2}\cr
&={1\over 2}\int{d^3p\over(2\pi)^3}\ln
{(1+\Pi_0)(p^2+p^2_3\Pi_0+{\vec p}^2\Pi_2)
+(\kappa-\Pi_1)^2\over p^2(1+\Pi_0|_{B\rightarrow 0})^2
+(\kappa-\Pi_1|_{B\rightarrow 0})^2}
+\Delta E_F+{B^2\over 2}\ ,\cr
}
\eqno(15)
$$
where $\Delta E_F$ and ${B^2/2}$ are contributions from the fermion and the
Maxwell term, respectively, the values of which are necessarily positive
in the limit of $\mu\rightarrow 0$. $D^{-1}_{\mu\nu}(p;B)$ is inverse of the
gauge field propagator with $B\ne 0$ defined by
$$
D^{-1}_{\mu\nu}(p;B)\equiv \Delta^{-1}_{\mu\nu}(p)
-\Pi_{\mu\nu}(p;\mu,m,B)\ ,
\eqno(16)
$$
and
$$
\eqalign{
\Delta^{-1}_{\mu\nu}(p)&=p^2[\delta_{\mu\nu}-(1-{1\over\xi}){p_\mu p_\nu\over
p^2}-\kappa{\epsilon_{\mu\nu\rho}p_\rho\over p^2}]\ ,\cr
\Pi_{\mu\nu}(p;\mu,m,B)
&=q^2\int{d^3k\over(2\pi)^3}
{\rm Tr}[\gamma_\mu S_F(k'+p)\gamma_\nu S_F(k')]\ .\cr
}
\eqno(17)
$$
Because of the existence of nonzero magnetic field, only the rotational
symmetry in two dimensional space survives. Taking account of the remaining
symmetry, and making regularization in
gauge invariant manner, we represent $\Pi_{\mu\nu}$ as
$$
\eqalign{
\Pi_{\mu\nu}(p;\mu,m,B)
&=(p_\mu p_\nu-p^2\delta_{\mu\nu})\Pi_0(p^2;\mu,m,B)\cr
&\quad-\epsilon_{\mu\nu\rho}p_\rho\Pi_1(p^2;\mu,m,B)\cr
&\quad+(1-\delta_{\mu 3})(1-\delta_{\nu 3})
 (p_\mu p_\nu-{\vec p}^2\delta_{\mu\nu})\cr
&\qquad\times[\Pi_2(p^2;\mu,m,B)-\Pi_0(p^2;\mu,m,B)]\ .\cr
}
\eqno(18)
$$
Using the propagator of eq.(9), $\Pi_i$ are found to be
$$
\eqalign{
\left[
\matrix{
\Pi_0 \cr \Pi_1 \cr \Pi_2
}
\right] &=
 {q^2 \over 4\pi^2}{1 \over |qB|}\int_{-\infty}^{\infty}dk_3
 \int_0^1 dw \int_0^{\infty} ds
 \left[
  \matrix{
  \pi_0 \cr \pi_1 \cr \pi_2
  }
 \right] \cr
&\times\exp{\biggl[-{s \over |qB|}\biggl\{(k_3-i\mu)^2+m^2+{1-w^2 \over 4}p_3^2
    +\biggl(
     {\cosh s -\cosh ws \over 2s\sinh s}
     \biggr)\vec{p}^2\biggr\}\biggr]},
}
\eqno(19)
$$
where
$$
\eqalign{
\pi_0 &= {s\cosh ws \over \sinh s}\biggl({1-w^2 \over 4}\biggr)+
        {s\cosh s \over \sinh^{3} s}
        \biggl({\cosh^2 s -\cosh^2 ws \over 4}\biggr),\cr
\pi_1 &= {s \over \sinh s}
        \biggl\{
        m\cosh ws -i(k_{3}-i\mu){qB \over |qB|}w\sinh ws
        \biggr\},\cr
\pi_2 &= {s\cosh s \over \sinh^{3} s}
        \biggl({\cosh^{2} s -\cosh^{2} ws \over 2}\biggr).
}
$$
In eq.(19), we cannot change the order of integrations
without losing the dependency of $\Pi_i$ on $\mu$.
In the asymptotic expansion with respect to $B$, $\Pi_i$ for small $B$ can
be written as
$$
\eqalign{
\Pi_{0,2}&=\theta(|m|-\mu)G(p^2;\mu=|m|,|m|)\cr
                  &\ +\theta(\mu-|m|)\theta(p^2-4(\mu^2-m^2))
                  G(p^2;\mu,|m|)+{\cal O}(B^2)\ ,\cr
\Pi_1\ \ &=\theta(|m|-\mu)F(p^2;\mu=|m|,|m|)\cr
                  &\ +\theta(\mu-|m|)\theta(p^2-4(\mu^2-m^2))
                    [F(p^2;\mu,|m|)\cr
                  &\ +qBH(p^2;\mu^2,|m|)]
                    +{\cal O}(B^2)\ ,\cr
}
\eqno(20)
$$
where
$$
\eqalign{
F(p^2;\mu,|m|)&={q^2\over2\pi}{m\over\sqrt{p^2}}
\tan^{-1}\bigl({\sqrt{p^2-4(\mu^2-m^2)}\over 2\mu}\bigr)\ ,\cr
G(p^2;\mu,|m|)&={q^2\over8\pi}{1\over(p^2)^{3/2}}
\Bigl[ 2\mu\sqrt{p^2-4(\mu^2-m^2)}\cr
&\qquad+(p^2-4m^2)\tan^{-1}\bigl({\sqrt{p^2-4(\mu^2-m^2)}\over 2\mu}\bigr)
\Bigr]\ ,\cr
H(p^2;\mu,|m|)&={q^2\over 4\pi}{1\over(p^2)^{3/2}}
                           \sqrt{p^2-4(\mu^2-m^2)}\ .\cr
}
\eqno(21)
$$
One finds from eq.(20) that if $|m|>\mu$, no terms proportional to
$B$ appear in $\Delta E$, since we get
$$
{\rm det}D^{-1}(p;B)={\rm det}D^{-1}(p;B\rightarrow 0)+{\cal O}(B^2)\ ,
\eqno(22)
$$
substituting eq.(20) into eq.(17). This fact means that $B\ne 0$ state with
the energy lower than the naive vacuum state does not exist in this region.

On the other hand, if $\mu>|m|$, then we get
$$
\eqalign{
{\rm det}D^{-1}(p;B)&=\theta(4(\mu^2-m^2)-p^2)\xi^{-1}(p^2)^2(p^2+\kappa^2)\cr
&\ +\theta(p^2-4(\mu^2-m^2))\xi^{-1}(p^2)^2\cr
&\ \times\{p^2(1+G(p^2;\mu,|m|))^2
+(\kappa- F(p^2;\mu,|m|))^2\cr
&\quad -2qB(\kappa- F(p^2;\mu,|m|))
H(p^2;\mu,|m|)\}+{\cal O}(B^2)\ .
}
\eqno(23)
$$
Thus, there exists a term proportional to $B$ in ${\rm det}D^{-1}(p;B)$, and
in $\Delta E$. Indeed, if we take a limit of $\mu\rightarrow 0$, keeping $\mu
>|m|$, then the energy difference eq.(15) becomes
$$
\Delta E |_{\mu\downarrow |m|\rightarrow 0}=
-{q^2\over 8\pi^3}\tan^{-1}({4\over\pi})\cdot qB+{\cal O}(B^{3/2})\ .
\eqno(24)
$$
The above result is just coincident with that of ref.[3]. Namely, there exists
a $B\ne 0$ state with lower energy than the naive vacuum, since the sign
of coefficient for linear $B$ term is necessarily negative. However, the most
important feature in the above analysis is that the $B\ne 0$ state, energy of
which is lower than that of naive vacuum state, can be constructed only when
the condition $\mu>|m|$ is kept, by hand.
This fact means that we need to supply, from the external source, the energy
which exceeds the lowest Landau level $m$ in order to induce the $B\ne 0$
state. Accordingly, we also have to take the massless limit $m\rightarrow 0$
so as to give rise to the spontaneous magnetization. Indeed, as shown in
eq.(24), the spontaneous magnetization really occurs in the limit $\mu
\downarrow |m|\rightarrow 0$.

To see this in more detail, let us consider the limit $\mu\rightarrow 0$ on
the $(\mu,m)$-plane (Fig.2). The $(\mu,m)$-plane is divided into two regions
by the
boundary line $\mu=|m|$. The necessary condition for $B\ne 0$ is given by
$\kappa=(q^2/2\pi)(m/|m|)$ in upper region $\mu<|m|$, while it is given by
$\kappa=(q^2/2\pi)(qB/|qB|)$ in the lower region $\mu>|m|$.

First, we consider the case of $m/|m|\ne qB/|qB|$. If $\kappa=(q^2/2\pi)
(m/|m|)$ is satisfied, the nontrivial vacuum with $B\ne 0$ is allowed only in
the upper region. The stable vacuum is, however, the naive vacuum with $B=0$
inside the upper region, and consequently, the spontaneous magnetization does
not appear. On the other hand, if $\kappa=(q^2/2\pi)(qB/|qB|)$ is satisfied,
the nontrivial vacuum is possible only in the lower region. We should therefore
take the limit $\mu\rightarrow 0$ only inside the lower region so as to make
the spontaneous magnetization possible. In the lower region, the vacuum with
$B\ne 0$ has energy lower than that of the naive vacuum, and if we take the
limit $\mu\downarrow |m|\rightarrow 0$, the spontaneous magnetization
actually occurs.

Next, let us consider the case of $m/|m|=qB/|qB|$ in which the necessary
condition for $B\ne 0$ is satisfied in the both of two regions, simultaneously.
In this case, we can take the limit $\mu\rightarrow 0$ for a fixed fermion
mass. It is possible that all of the points on the $(\mu,m)$-plane approach
the $m$-axis along the horizontal lines which cross over the boundary line
$\mu=|m|$. However, since those points necessarily go to inside the upper
region where only the naive vacuum is stable, the spontaneous magnetization
cannot be realized. Therefore, we should take the limit $\mu\downarrow |m|
\rightarrow 0$, even if the relation $m/|m|=qB/|qB|$ is satisfied.

In any case, in order to realize the spontaneous magnetization, we need to
take the massless limit $|m|\rightarrow 0$ as well as $\mu\rightarrow 0$,
keeping fermion mass $|m|$ lower than the chemical potential $\mu$.
In ref.[3], the nontrivial vacuum was obtained in the
massless limit of fermions. This fact is consistent with the above result.
It is important to emphasize, however, that
if the chemical potential was not introduced, we could not find out
that the massless fermion limit is a necessary condition to realize the
spontaneous magnetization.

So far, we have dealt with the system with the originally massive fermion.
The spontaneous magnetization occurs at massless limit of the fermion.
It is instructive to consider the case of originally massless fermion. In
this case, however, the infra-red behavior of vertex functions breakes the
non-renormalization theorem of Coleman-Hill, which is extended to the finite
density system. We therefore cannot derive the definite result about the
charge condensation only from the one loop calculation. There is a possibility,
however, that the originally massless fermion gains its mass dynamically.
So, it is necessary to examine the dynamical generation of
fermion	mass as well as the charge condensation.

Recently, it was shown by Kondo et al.$^{[7]}$ that fermion mass is
dynamically generated in QED$_3$ with a Chern-Simons term, which is just
the same model as one treated here. In their picture, we should regard
the fermion mass $m$, which appears in the lagrangian, as an external source
for the condensation $\langle\bar{\psi}\psi\rangle$. On the other hand,
the chemical potential was taken to zero from the beginning, and so
the vacua are aligned on the $m$-axis in $(\mu,m)$-plane. They found that
the condensation $\langle\bar{\psi}\psi\rangle$ remains nonzero in the limit
$|m|\rightarrow 0$ along the $m$-axis. As for the spontaneous magnetization,
only the naive vacuum with $B = 0$ is realized within their procedure.
Thus, in order to understand the vacuum structure completely, it becomes the
main problem to clarify whether or not both the fermion
mass generation and the spontaneous magnetization occur, simultaneously.

Finally, we make some comments about the Nambu - Goldstone theorem.
Eq.(12) and eq.(23) imply that the photon becomes massless$^{[2]}$.
It was pointed out in ref.[3] that the massless photon can be interpreted
as the Nambu-Goldstone (NG) boson for the breakdown of Lorentz invariance.
In order to identify the massless photon with the NG boson,
we have to show that the continuous symmetry, which is the Lorentz invariance
in the Minkowski space-time, is spontaneously broken.
It is not clear within the analysis in ref.[3], however, whether
the occurrence of magnetization is really spontaneous or not,
since this cannot be confirmed without introducing the chemical
potential as an external source for $\langle\psi^\dagger\psi\rangle$.
On the other hand, we have explicitly shown that the spontaneous
magnetization occurs in the limit $\mu\downarrow|m|\rightarrow 0$ and thereby
the $SO(3)$ symmetry of the Euclidean space-time is surely broken
spontaneously. Thus, in our case, we can make sure that the massless
photon is identified with the NG boson for the $SO(3)$ symmetry breaking.

In conclusion, we have shown in the explicit manner that the charge
condensation actually occurs ``spontaneously", using the chemical potential.
The spontaneous magnetization, however, is not the general feature of the
system, but can occur when $\mu$ approaches zero along the peculiar
direction $\mu \downarrow|m|\rightarrow 0$ on $(\mu,m)$-plane.
This result suggests that there is an implicit connection between the
spontaneous magnetization and the fermion mass generation, which remains as
a future problem.
\vskip 1cm
\ack
We would like to thank Prof. K. Yamawaki and Prof. S. Kitakado for useful
comments and reading the manuscript. We are grateful to K.-I. Kondo and A.
Shibata for helpful suggestions and valuable discussions.
We would like to thank S. Tanimura, H. Otsu and H. Nakatani for encouraging
suggestions at early stage on this work.
\vfill
\eject
\centerline{\fourteenrm REFERENCES}
\item{[1]}{R. Prange and S. Girvin, eds. {\it The Quantum Hall Effect},
(Springer-Verlag, 1987); F. Wilczek, {\it Fractional Quantum Statistics
and Anyon Superconductivity}, (World Scientific, 1990); E. Fradkin, {\it
Field Theories of Condensed Matter System}, (Addison-Wesley, 1991).}
\item{[2]}{J. D. Lykken, J. Sonnenschein and N. Weiss, Int. J. Mod. Phys.
{\bf A6} (1991) 5155.}
\item{[3]}{Y. Hosotani, Phys. Lett. {\bf B319} (1993) 332; Phys. Rev. {\bf
D51} (1995) 2022; UMN-TH -1304/94 {\tt (hep-th /9407188)}; UMN-TH -1308/94 {\tt
(hep-th /9408148)}; S. Kanemura and T. Matsushita, OU-HET 212
{\tt(hep-th /9505146).}}
\item{[4]}{R. Jackiw, Phys. Rev. {\bf D9} (1974) 1688.}
\item{[5]}{A. N. Redlich, Phys. Rev. Lett. {\bf 52} (1984) 18; Phys. Rev. {\bf
D29} (1984) 2366.}
\item{[6]}{S. Coleman and B. Hill, Phys. Lett. {\bf 159B} (1985) 184.}
\item{[7]}{K.-I. Kondo, T. Ebihara, T. Iizuka and E. Tanaka, CHIBA-EP-77-REV
{\tt(hep-th /9404361)}; K.-I. Kondo and P. Maris, Phys. Rev. Lett. {\bf 74}
(1995) 18; CHIBA-EP-85 /DPNU-94-51 ({\tt hep-th /9501280}); K.-I. Kondo,
CHIBA-EP-89.}
\vfill
\eject
\centerline{\fourteenrm FIGURE CAPTIONS}
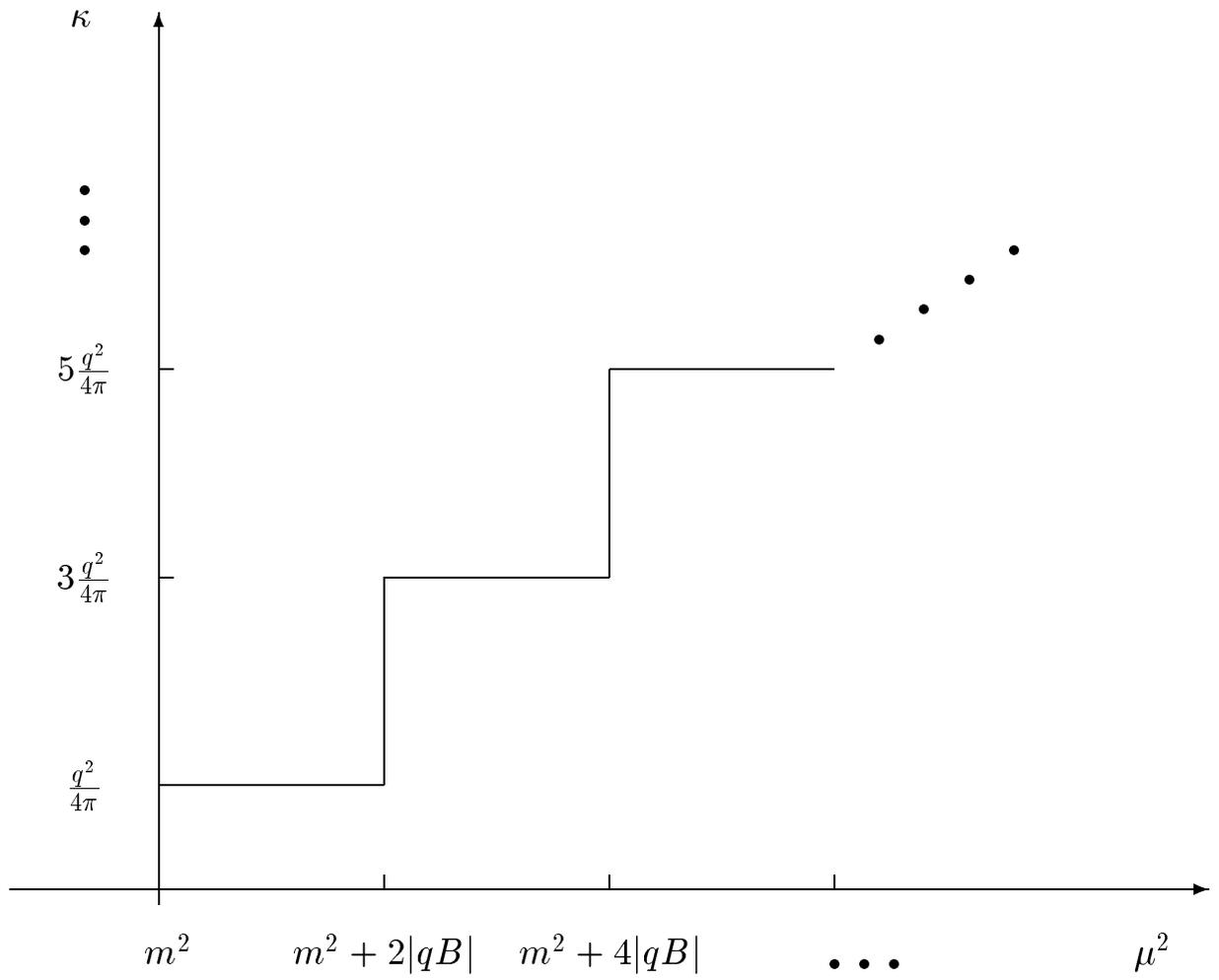
\item{Fig.1}{The necessary condition for $B\ne 0$.}
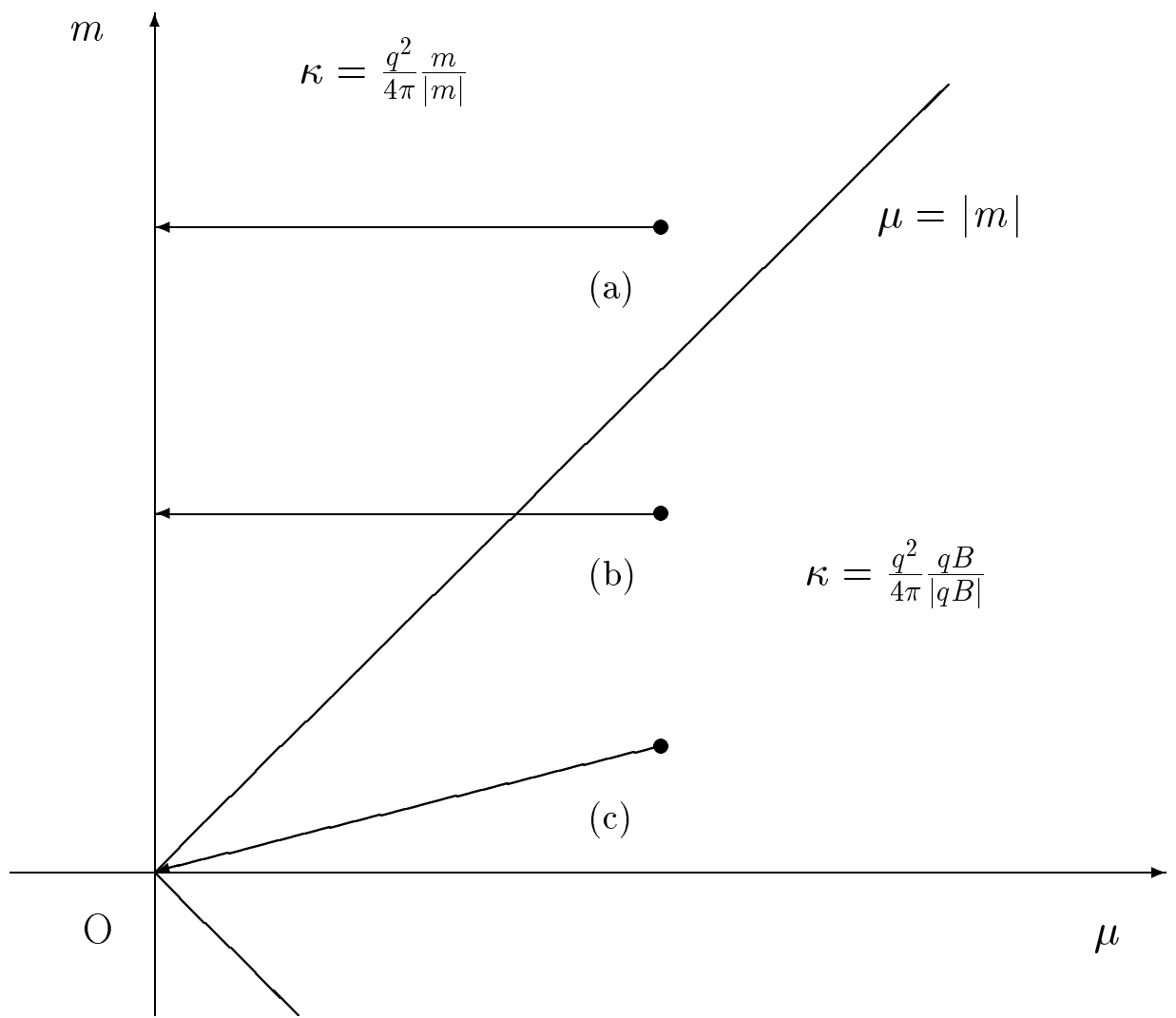
\item{Fig.2}{Possibilities of $\mu\rightarrow 0$ limit, keeping the necessary
condition for $B\ne 0$. When $m/|m|\ne qB/|qB|$, (a) is allowed if $\kappa=
(q^2/4\pi)(m/|m|)$, while (c) is allowed if $\kappa=(q^2/4\pi)(qB/|qB|)$.
When $m/|m|=qB/|qB|$, (a), (b) and (c) are allowed if $\kappa=(q^2/4\pi)
(qB/|qB|)$.}
\end